\documentclass[twocolums,showpacs,preprintnumbers,amsmath,amssymb]{revtex4}
\usepackage[usenames]{color}
\usepackage{graphicx}
\usepackage{epsfig}
\usepackage{graphicx}
\usepackage{dcolumn}
\usepackage{bm}
\begin{document}
\title{Derivation of Dirac, Klein-Gordon, Schrodinger, Diffusion and quantum heat transport equations from a universal quantum  wave equation}
\author{Arbab I. Arbab\footnote{Email: aiarbab@uofk.edu}}
\affiliation{Department of Physics, Faculty of Science, University of Khartoum, P.O. Box 321, Khartoum
11115, Sudan,}
\begin{abstract}
A universal quantum wave  equation that yields Dirac,
Klein-Gordon, Schrodinger and quantum heat equations is derived.
These equations are related by complex transformation of space,
time and mass. The new symmetry exhibited by these equations is
investigated. The universal quantum equation yields Dirac equation in two
ways: firstly by replacing the particle my $m_0$ by $im_0$, and
secondly by changing space and time coordinates by $it$ and $i\vec{r}$, respectively.
\end{abstract}
\pacs{03.65.Ta, 03.65.-w, 03.65.Pm, 02.30.Jr, 03.50.Kk}
\maketitle
\section{Introduction}
Adler and others have developed the quaternionic quantum mechanics [1]. However, we have recently adopted a different view and came up with a generalizes Klein-Gordon equation [2]. Moreover, the Dirac equation can be obtained from this equation by either replacing the time and space coordinates by their imaginary ones. Or by replacing the mass of the particle $m_0$ by the imaginary mass $im_0$.
Moreover, this equation exhibits the invariance of the transformation of space-time and mass
by their imaginary ones.   The universal quantum wave equation we have derived here  is the relativistic quantum heat
transport equation of the equation proposed recently  by Kozlowski and  Marciak-Kozlowska [3].  It is also invariant under the
combined space-time and mass imaginary transformations, viz.,
$t\rightarrow i\, t$, $\vec{r}\rightarrow i\, \vec{r}$, and $m_0\rightarrow - i\, m_0$,. Dirac, Klein-Gordon and  Schrodinger equations are invariant under this complex transformation. t'Hooft and Nobbenhuis have recently argued that the invariance under space-time complex transformation results in a vacuum state with strictly vanishing vacuum energy [4]. However, ref. [4] have left the mass of the particle unchanged, a fact that affected the integrity and the  invariance of the other physical quantity under this transformation. In as much as energy and mass are intimately related, so that a transformation of the former would transform the latter. Allowing this transformation will retain the whole physics of the system. This property is investigated in this work.
\section{Quaternionic quantum mechanics: Dirac-like equation}

Consider a particle described by the quaternion wavefunction
$\Psi=(\frac{i}{c}\psi_0\,, \vec{\psi})$. This is equivalent to
spinor representation of ordinary quantum mechanics that we have
recently proposed [1]. The evolution of this quaternion
wavefunction is defined by the three equations [1]
\begin{equation}
\vec{\nabla}\cdot\vec{\psi}-\frac{1}{c^2}\frac{\partial \psi_0}{\partial t}-\frac{m_0}{\hbar}\,\psi_0=0\,,
\end{equation}
\begin{equation}
\vec{\nabla}\psi_0-\frac{\partial \vec{\psi}}{\partial t}-\frac{m_0c^2}{\hbar }\,\vec{\psi}=0\,,
\end{equation}
and
\begin{equation}
\vec{\nabla}\times\vec{\psi}=0\,.
\end{equation}
Equations (1) - (3) yield the two wave equations
\begin{equation}
\frac{1}{c^2}\frac{\partial^2\vec{\psi}}{\partial t^2}-\nabla^2\vec{\psi}+2\left(\frac{m_0}{\hbar}\right)\frac{\partial\vec{\psi}}{\partial t}+\left(\frac{m_0c}{\hbar}\right)^2\vec{\psi}=0\,,
\end{equation}
and
\begin{equation}
\frac{1}{c^2}\frac{\partial^2\psi_0}{\partial t^2}-\nabla^2\psi_0+2\left(\frac{m_0}{\hbar}\right)\frac{\partial\psi_0}{\partial t}+\left(\frac{m_0c}{\hbar}\right)^2\psi_0=0\,.
\end{equation}
An equation of similar form is known as \emph{the generalized equation of telegraphy} [5]. Hence, Eq.(4) represents a generalized telegraph equation.
This generalized wave  is written as
\begin{equation}
\frac{1}{c^2}\left(\frac{\partial^2\phi}{\partial t^2}+p\frac{\partial \phi}{\partial t}+q \,\phi\right)=\nabla^2\phi\,,
\end{equation}
where $p$ and $q$ are constant.
The solution of Eq.(4) will be of the form
\begin{equation}
\psi=A\exp(-\frac{m_0c^2}{\hbar}t)\exp(\pm 2\pi\,i(ckt-\vec{k}\cdot \vec{r}))\,,\,\,\, A=\rm const.\,.
\end{equation}
It then follows that this wave attenuates with time. This wavefunction represents a wave that travels without distortion. Such a wave is said to be relatively undistorted wave. It is of great importance in telephone line construction, where the signal will be attenuated  as it propagates but not distorted. This may probably mimic the propagation of heat wave. However, if Eq.(4) is an equation of a fundamental particle, then the particle shape doesn't distort as it travels. But it attenuates due to its mass.

Equations (1) - (3) are invariant under the complex transformation
\begin{equation}
t\rightarrow i\, t\,\,, \qquad \vec{r}\qquad \rightarrow i\, \vec{r} \,\,, \qquad m_0\rightarrow -i\, m_0\,.
\end{equation}
We call here the complex transformation in Eq.(7) the \emph{universal complex transformation}-UCT. Under the UCT, the energy and momentum will transform as
\begin{equation}
E\rightarrow - i\, E\,\,, \qquad \vec{p}\rightarrow - i\, \vec{p} \,,
\end{equation}
so that  the phase angle ($\phi$) of the wave function  will
transform as
\begin{equation}
\phi=Et-\vec{p}\cdot \vec{r}\qquad \rightarrow \qquad \phi\,.
\end{equation}
Therefore, this transformation does not alter the phase factor. The angular momentum ($\vec{L}$), power ($P$) and force ($\vec{F}$) transform as follows
\begin{equation}
\vec{L}\rightarrow \vec{L}\,\,, \qquad P\rightarrow - P \,,\qquad \vec{F}\rightarrow - \vec{F} \,.
\end{equation}
Maxwell equations are invariant under the UCT if the electric and magnetic fields, $\vec{E}$ and $\vec{B}$, transform as
\begin{equation}
\vec{E}\rightarrow -\vec{E}\,\,, \qquad \vec{B}\rightarrow - \vec{B} \,,
\end{equation}
so that the vector and scalar fields, $\vec{A}$ and $\varphi$, transform as
\begin{equation}
\vec{A}\rightarrow -i\, \vec{A} \,, \qquad \varphi\rightarrow -i \,\varphi\,.
\end{equation}
The electric current and charge densities, $\vec{J}_e$ and $\rho_e$, transform as\footnote{The charge ($q$), voltage ($V$), current ($I$), capacitance ($C$) and inductance ($L$) transform as: $ q\rightarrow - i\, q\,,\,V\rightarrow - i\, V\,,\,I\rightarrow - i\,I\,,\, C\rightarrow  i\,C\,,\, L\rightarrow  i\,L $.}
\begin{equation}
\vec{J}_e\rightarrow i\, \vec{J}_e \,, \qquad \rho_e\rightarrow i \,\rho_e\,.
\end{equation}
Moreover, owing to UCT, the permittivity, permeability, Planck's constant and Newton's constant transform as
\begin{equation}
\varepsilon_0\rightarrow \varepsilon_0\,  \,, \qquad \mu_0\rightarrow \mu_0\,, \qquad G\rightarrow -G\,,\qquad\hbar\rightarrow \hbar\,.
\end{equation}
We, according, believe that the UCT must be a symmetry of nature. Thus, the physical world must be invariant under the UCT. Recently, t'Hooft and Nobbenhuis have  considered a complex invariance that results in a vacuum state with strictly vanishing vacuum energy. They have argued that the vacuum state is the only unique state that is invariant under this transformation and all other states break this symmetry [4].

In the context of gravitation, the invariance of Einstein field equations under UCT requires that [6]
\begin{equation}
R_{\mu\nu}\rightarrow -\, R_{\mu\nu} \,, \qquad \rho_m\rightarrow \,\rho_m\,,\qquad G\rightarrow -\,G \,,
\end{equation}
where $\rho_m$ is the matter density. The force transformation in Eq.(11) implies that in the complex world forces are inverted. However, since gravity is always attractive in the real world, a negative $G$ guarantees that the gravitational force between two masses in the complex world would still be attractive.
\section{The quantum heat transport equation}
More recently,  Kozlowski and  Marciak-Kozlowska have derived a new quantum heat transport equation [3]. According to their equation
the temperature, $T$, satisfies the wave equation
\begin{equation}
\frac{1}{v^2}\frac{\partial^2T}{\partial t^2}-\nabla^2T+\left(\frac{m}{\hbar}\right)\frac{\partial T}{\partial t}+\frac{2Vm}{\hbar^2}\,T=0\,,
\end{equation}
where $m$ is the mass of the heat carriers and $V$ is the
nonthermal potential. They found that for undistorted thermal
wave, i.e., a wave which preserves the shape in the field of the
potential $V$, the relation
\begin{equation}
V\,\tau=\frac{\hbar}{8}\,,\qquad \tau=\frac{\hbar}{mv^2}\,.
\end{equation}
holds. If we compare Eq.(17) with Eq.(4), we obtain
\begin{equation}
m=2m_0\,, \qquad V=\frac{1}{4}\, m_0c^2\,.
\end{equation}
Hence, our equation, Eq.(4),  represents the quantum relativistic version of Eq.(17). Equation (4) can also represent the equation of an undistorted particle (wave) that is attenuated as it propagates in space-time due to its mass (inertia).

\section{Ordinary quantum mechanics: Dirac equation}
Dirac's equation can be written in the form [7]
\begin{equation}\label{1}
\frac{1}{c}\frac{\partial\psi}{\partial t}+\vec{\alpha}\cdot\vec{\nabla}\psi+\frac{im_0c\,\beta}{\hbar}\psi=0\,.
\end{equation}
where   $\beta=\left (\begin{array}{cc}
  1 & 0 \\
  0 & -1 \\
\end{array}\right),$    $\alpha=\left (\begin{array}{cc}
 0 & \vec{\sigma}   \\
   \vec{\sigma}  & 0 \\
\end{array}\right),$ $\alpha^2=\beta^2=1$ and $\vec{\sigma}$ are the Pauli matrices. We have found that this equation can be written as
\begin{equation}
\frac{1}{c^2}\frac{\partial^2\psi}{\partial t^2}-\nabla^2\psi+2\left(\frac{m_0i\beta}{\hbar}\right)\frac{\partial\psi}{\partial t}-\left(\frac{m_0c}{\hbar}\right)^2\psi=0\,,
\end{equation}
and
\begin{equation}
\frac{1}{c^2}\frac{\partial^2\psi}{\partial t^2}-\nabla^2\psi-2\left(\frac{m_0\,i}{\hbar}\right)\beta\, c\,\vec{\alpha}\cdot\vec{\nabla}\psi+\left(\frac{m_0c}{\hbar}\right)^2\psi=0\,.
\end{equation}
Equation (21) can be split into positive and negative energy equations respectively
\begin{equation}
\frac{1}{c^2}\frac{\partial^2\chi}{\partial t^2}-\nabla^2\chi+2\left(\frac{m_0i}{\hbar}\right)\frac{\partial\chi}{\partial t}-\left(\frac{m_0c}{\hbar}\right)^2\chi=0\,,
\end{equation}
and
\begin{equation}
\frac{1}{c^2}\frac{\partial^2\varphi}{\partial t^2}-\nabla^2\varphi-2\left(\frac{m_0i}{\hbar}\right)\frac{\partial\varphi}{\partial t}-\left(\frac{m_0c}{\hbar}\right)^2\varphi=0\,.
\end{equation}
Equations (23) and (24) can be obtained from Eq.(4) by replacing the mass $m_0$ by $\pm\, i\,m_0$, respectively.
They can be compared with the Klein-Gordon equation of  spin - 0 particles, i.e.,
\begin{equation}
\frac{1}{c^2}\frac{\partial^2\psi}{\partial t^2}-\nabla^2\psi+\left(\frac{m_0c}{\hbar}\right)^2\psi=0\,.
\end{equation}
If the dissipation term in Eq.(4) can be neglected, the resulting
equation is the Klein-Gordon equation, Eq.(25). Equation (4) tells us that the particles moves as if it were massless and then  moves as a free particle as governed by Klein-Gordon equation. One can estimate the characteristic time during which it moves as a massless particle as $t_c=\frac{\hbar}{m_0c^2}$. For a pion ($\pi^\pm$), one finds $t_c=4.69\times 10^{-19}\rm s$. With attosecond spectroscopy one can detect this motion. Therefore, a pion moves as a massless particle before this time. Generally heavier particles free themselves after a shorter time. Such a time can be measured experimentally from the motion of spin-0 particles and compared with the theoretical findings. The cause of this initial resistance is the inertia of the particle that the wave describes. However, in de Broglie theory the motion of the particle is not determined. But de Broglie associates a wave with a particle motion. Can we say that the wave has an \emph{inertial} property? It is interesting to mention here  that spin - 1/2 particles move without attenuation, as evident from Eq.(7).

Hence, Eq.(4) represents a generalized Klein-Gordon equation. Equation (24) can
be written in the form
\begin{equation}
i\hbar\frac{\partial\varphi}{\partial t}=-\frac{\hbar^2}{2m_0}\nabla^2\varphi+\frac{\hbar^2}{2m_0c^2}\frac{\partial^2\varphi}{\partial t^2}-\frac{m_0c^2}{2}\,\varphi\,,
\end{equation}
which can be seen as a generalized Schrodinger's equation.
Equation (4)  can be written as
\begin{equation}
\frac{\partial\psi}{\partial t}=\frac{\hbar}{2m_0}\nabla^2\psi-\frac{\hbar}{2m_0c^2}\frac{\partial^2\psi}{\partial t^2}-\frac{m_0c^2}{2\hbar}\,\psi\,.
\end{equation}
This reduces to the diffusion equation, when we neglect the last two terms, viz.,
\begin{equation}
\frac{\partial\psi}{\partial t}=D\nabla^2\psi\,,
\end{equation}
where the diffusion coefficient, $D=\frac{\hbar}{2m_0}$.

It is interesting to know that Dirac, Klein-Gordon, Schrodinger equations are invariant under the UCT. Using Eq.(8), the commutator algebra is covariant under the UCT, viz.,
\begin{equation}
\left[ x_i\,, p_j\right]=i\hbar\, \delta_{ij}\,,\qquad \left[ L_i\,, L_j\right]=i\hbar\, L_k\, \epsilon_{ijk}\,.
\end{equation}
Applying the complex transformation
\begin{equation}
t\rightarrow i\, t\,,\qquad \qquad \vec{r}\rightarrow i\, \vec{r}\,,
\end{equation}
in Eq.(4) yields the Dirac positive energy equation, i.e.,
Eq.(23). This implies that Eq.(4) is relativistically invariant.
Thus, Eq.(4) is equivalent to Dirac equation in imaginary space
and time. The coordinates transformation in Eq.(30) is equivalent
to rotation of the space and time coordinates by an angle of 90
degrees. It is also equivalent, in the relativity theory, to replacing the squared interval $ds^2$ by $-ds^2$ .
\section{Schrodinger equation}
Now apply the transformation in Eq.(30) to Eq.(1) to obtain
\begin{equation}
\vec{\nabla}\cdot\vec{\psi}-\frac{1}{c^2}\frac{\partial \psi_0}{\partial t}-\frac{m_0i}{\hbar}\,\psi_0=0\,.
\end{equation}
Equation (3) can be satisfied by choosing
\begin{equation}
\vec{\psi}=a \vec{\nabla}\psi_0\,,\qquad a=\rm const.\,\,\,\,.
\end{equation}
Substituting Eq.(32) in Eq.(31) yields
\begin{equation}
i\hbar\frac{\partial \psi_0}{\partial t}=\frac{-\hbar^2}{2m_0}\,
\nabla^2 \psi_0+m_0c^2\psi_0\, \,,
\end{equation}
where $ a=\frac{i\hbar}{2m_0c^2}\,.$ Equation (33) is the
Schrodinger equation for a particle with potential equals to the
rest mass energy of the particle, i.e., $V=m_0c^2$. It can also be
written as
\begin{equation}
H\psi_0=E\psi_0\,\,,\qquad {\rm where}\qquad  E=E_K+m_0c^2 \,,
\end{equation}
where $E_K=\frac{p^2}{2m_0}$ is the nonrelativistic kinetic energy
of the particle.
Equation is thus the nonrelativistic approximation of the equation
\begin{equation}
H\psi=E\psi\,,
\end{equation}
where
\begin{equation}
E=\sqrt{p^2c^2+m_0^2c^4}\,.
\end{equation}
Notice that unlike  Klein-Gordon equation which
yields Schrodinger equation in the nonrelativistic limit, Dirac
equation yields Weyl equation instead. However, we have found here
that the universal quantum wave equation yields Schrodinger
equation. Alternatively, we can say that Eq.(33) represents the motion of a free particle. Such an equation should replace the ordinary Schrodinger equation for a free particle which doesn't involve the rest mass energy term.
\section{Concluding Remarks}
We have derived a new wave equation that retains Schrodinger, Dirac, Klein-Gordon, diffusion and quantum heat transport equations for certain transformations and approximations. This equation is relativistically invariant.  We have found that the UCT is a true law of nature where all equations of motion must satisfy. The full physical meaning of the UCT will be our future endeavor.
\section*{Acknowledgement}
I would like to thank H.M. Widatallah for enlightening and constructive comments.
\section*{References}
\hspace{-0.35cm}$[1]$ Adler, S. L.,  \emph{Quaternionic Quantum Mechanics and Quantum Fields}, Oxford University Press, New York, (1995).;
David Finkelstein1, Josef M. Jauch, Samuel Schiminovich, and David Speiser., \emph{Foundations of Quaternion Quantum Mechanics}, J. Math. Phys. 3, 207 (1962).\\
$[2]$ Arbab, A. I.,  \emph{The quaternionic quantum mechanics}, eprint arXiv:1003.0075.\\
$[3]$ Miroslaw Kozlowski and Janina Marciak-Kozlowska, arXiv:q-bio/0501031v2.\\
$[4]$ Gerard 't Hooft, Stefan Nobbenhuis, \emph{Invariance under complex transformations, and its
relevance to the cosmological constant problem},  Class. Quantum Grav. 23, 3819 (2006).
\\
$[5]$ Coulson, C. A.,  \emph{Waves}, Logman group Ltd., pp. 17 (1977).\\
$[6]$ Weinberg, S., \emph{Gravitation and cosmology}, John Wiley \& Sons, Inc. New York, (1972).\\
$[7]$ Bjorken, J. D., and Drell, S. D., \emph{Relativistic Quantum Mechanics}, McGraw-Hill, (1964).\\

\end{document}